\documentclass[aps,prl,superscriptaddress]{revtex4}
\usepackage[cp866]{inputenc}
\usepackage[english]{babel}
\usepackage{epsfig,graphicx}
\usepackage{indentfirst}
\usepackage{amsfonts,amssymb,latexsym,amsmath,enumerate,amsthm}
\usepackage{bm}
\usepackage{color}

\graphicspath{{FIGURES/}}

\newcommand{\be}{\begin{equation}}
\newcommand{\ee}{\end{equation}}
\newcommand{\bea}{\begin{eqnarray}}
\newcommand{\eea}{\end{eqnarray}}
\newcommand{\dst}{\displaystyle}
\newcommand{\fr}[2]{\frac{{\dst #1}}{{\dst #2}}}

\usepackage{array}

 \newcommand{\nn}{\nonumber}

\newcommand{\bp}{{\bf p}}
\newcommand{\br}{{\bf r}}
\newcommand{\bn}{{\bf n}}

\newcommand{\bb}{{\bf b}}
\newcommand{\bE}{{\bf e}}

 \newcommand{\bB}{{\bf B}}
 
\newcommand{\bee}{{\rm e}}
\newcommand{\bii}{{\rm i}}

\begin{document}

\title{Elastic scattering of twisted neutrons by nuclei}
\vspace*{15px}

\author{A.\,V.~Afanasev}
\affiliation{Department of Physics,
The George Washington University, Washington, DC 20052, USA}

\author{D.\,V.~Karlovets}
\affiliation{Tomsk State University, Lenina Ave.36, 634050 Tomsk, Russia}

\author{V.\,G.~Serbo}
\affiliation{Novosibirsk State University, RUS-630090, Novosibirsk, Russia}
\affiliation{Sobolev Institute of Mathematics, RUS-630090, Novosibirsk, Russia}
\date{February 18, 2021}

\begin{abstract}
We present a theoretical formalism for scattering of the twisted neutrons by nuclei in a kinematic regime where interference between the Coulomb interaction and the strong interaction is essential. Twisted neutrons have definite quantized values of an angular momentum projection along the direction of propagation, and we show that it results in novel observable effects for the scattering cross section, spin asymmetries and polarization of the scattered neutrons. We demonstrate that additional capabilities provided by beam's orbital angular momentum enable new techniques for measuring both real and imaginary parts of the scattering amplitude. Several possible observables are considered, for which the targets may be either well-localized with respect to the spatial beam profile, or the scattering occurs incoherently on nuclei in a bulk target. The developed approach can be applied  to other nuclear reactions with strongly interacting twisted particles.
\end{abstract}
\maketitle

\section{ 1. Introduction}

Neutron scattering and interferometry provide fundamental probes of electromagnetic, strong, weak interactions and gravity \cite{Byrne_book,Rauch}. State-of-the-art reactor or spallation neutron facilities are currently operational around the world \cite{facilities}, and new facilities come online \cite{SNS} or are under construction \cite{ESS}. Recently \cite{exp-2015}, a new important capability for the fundamental neutron science was added: namely, the thermal (0.27nm) neutron beams were formed with a nonzero projection of the orbital angular momentum -- referred to as twisted neutrons --  using a method of spiral phase plates. Other novel approaches for generating the beams of twisted neutrons \cite{exp-2018,exp-2019,Geerits20} can also be applied to ultra-cold neutrons with sub-micron wave lengths.

Preparation of beams of light and matter in a pre-defined quantum state of angular momentum projection on beam's direction was previously achieved for photons and electrons \cite{Allen,Mono,Bliokh17,UFN}. These beams open new directions for controlling quantum states of matter, for optical and electron microscopy, quantum information, quantum communications and quantum computing. In the analysis of quantum amplitudes, the twisted-electron scattering allows one to access a Coulomb phase \cite{Ivanov_Phase, Karlovets_Phase}. Photo-excitation of atoms by the twisted photons provides approaches for separation of the transition amplitudes into different multipoles \cite{Afanasev-PRA18,Schultz-19}.
In the domain of elementary particle physics, new effects in the collisions of twisted beams were pointed out in Refs.~\cite{Ivanov-PRL20,Ivanov-PRD20}. Nuclear reactions caused by twisted gamma-rays and twisted neutrons were studied theoretically in Ref.~\cite{Afanasev-JPG18}. The use of twisted neutrons can become a part of multimode-entangled neutron interferometry discussed in Ref.\cite{Lu20} that would identify quantum-entangled degrees of freedom in matter.

In this paper, we analyze what novel information about the scattering amplitude can be learned from the twisted neutrons' being elastically scattered on a zero-spin atomic nucleus. A special case of the Schwinger scattering is considered \cite{Schwinger-48,Shull}, for which the interference of electromagnetic and strong interaction results in a characteristic spin asymmetry. Since a twisted neutron beam represents a (partially) coherent wave packet, different scattering scenarios lead to different observable effects. After a brief review of the scattering formalism for standard, plane-wave, neutron beams (Section 2), we develop a formalism of twisted-neutron scattering for macroscopic targets (Section 3), for which the cross section is an incoherent sum of cross sections for individual nuclei. We find that the angular dependence of the cross section is altered for the twisted particles, while the absorptive part of the amplitude is responsible for transverse spin asymmetry, as in a non-twisted case. Twisted neutron beams may be prepared in states of superposition of several angular momenta; we show that in this case scattering off macroscopic targets develops dependence on the longitudinal component of neutron spin. This is in stark contrast with conventional, non-twisted, neutron scattering, for which such a spin asymmetry is forbidden by parity conservation. Further we consider scattering on a single nucleus with a fixed  transverse position with respect to beam's axis (Section 4) and demonstrate that under this condition the scattering spin asymmetry is due to both longitudinal and transverse spin and, in addition, spin asymmetries have contributions from both real and absorptive parts of the nuclear amplitude. These features of the spin asymmetries still hold for mesoscopic targets, as we show in Section 5.

In summary, the magnitude of the predicted new effects depends on the parameters of the twisted neutron beams and experimental approaches: for well-localized targets and/or high angular resolution setup we demonstrate possible spin asymmetries in tens of per cent, while for bulk targets after the averaging over the nuclei positions, the effects may reduce to $10^{-6}$ levels. It should be noted that existing experimental programs studying parity-violation effects aim to measure even smaller asymmetries of 10$^{-8}$ \cite{Gericke08,Musolf-review}, therefore we believe that measurements of the predicted spin effects from the twisted neutrons are feasible and even necessary for separation of parity-conserving and parity-violating mechanisms of strong interactions.

This paper is a substantially expanded version of a Rapid Communication \cite{Afanasev-PRC19}.

\section{2. The standard case of plane-wave neutrons}

Here, we briefly review the formalism of neutron scattering on a spin-zero nucleus in the Schwinger regime \cite{Schwinger-48}, $i.e.$, when both Coulomb interaction and the strong interaction are essential. Let the initial neutron, prior to approaching the target, be in a plane-wave state with a momentum $\bp$ and a wave function $w\,\bee^{\bii \bp \br/\hbar}$, where the spinor $w= w^{(\lambda)}(\bn)$ with a helicity $\lambda$ is normalized as $w^\dag w=1$. The final neutron's wave function is $w'\,\bee^{\bii \bp' \br/\hbar}$. We neglect the target recoil, so that $p=p'$, and introduce the unit vectors $\bn=\bp/p$ and $\bn'=\bp'/p$ with the spherical angles $\theta, \,\varphi$ and $\theta', \,\varphi'$. The corresponding scattering amplitude is (see, for example, Ref.~\cite{BLP}, Sec. 42)
 \bea
 \label{1}
&& f_{\lambda \lambda'}(\bn,\,\bn')= w_{\lambda'}^{'\dag} \left(a+\bii {\bB} \bm \sigma\right) w_\lambda,
  \, \,
\bB = \beta\, \fr{\bn \times \bn'}{(\bn - \bn')^2},\nonumber \\
 &&\beta= \fr{\mu_n Z e^2}{m_p c^2}=-Z\times\,2.94\times 10^{-16}\;\mbox{cm},
 \eea
were $\bm \sigma$ are the Pauli matrices describing the neutron spin $\hat {\bf s}= \frac 12\, \bm \sigma$, $\mu_n=-1.91$ (in nuclear magnetons) and $m_p$ is the proton mass. Here,  $a$ is the nuclear amplitude while $\bii {\bB} \bm \sigma$ relates to the electromagnetic interaction of the neutron's anomalous magnetic moment with a nucleus.  Interference of these amplitudes in the cross section allows for important measurements of a phase of the nuclear amplitude. For thermal neutrons with the energies near 25 meV and an $^{197}_{79}\rm Au$ nuclear target ($a$=7.63 fm \cite{NISTdata}), the relevant  parameters are
 \be
 \varepsilon \equiv |\beta/a| \approx 0.03, \; |(\mbox{Im}\,a)/a| \approx 2\times10^{-4}.
 \label{parameters-for-Au}
 \ee

The standard cross section {\it summed over spin states of final neutrons} has the form
 \be
 \fr{d\sigma^{\rm(st)}(\bn,\,\bn',\bm \zeta)}{d\Omega'}= \sum_{\lambda'}\left|f_{\lambda \lambda'}(\bn,\,\bn')\right|^2=
 |a|^2+|\bB|^2 + 2  \, (\bB \bm \zeta)\,\mbox{Im}\, a,
 \label{crspl}
 \ee
where $\bm \zeta=(\bm \zeta_\perp, \zeta_z)$ is the polarization of the initial neutron beam, $|\bm \zeta| \leq 1$.
Assuming that the vector $\bn$ is directed along the $z$ axis ($i.e.$, that $\bn={\bf e}_z=(0,\,0,\,1)$), we find

\bea
 &&\fr{d\sigma^{\rm(st)}({\bf e}_z,\,\bn', \bm \zeta)}{d\Omega'} \, =
 |a|^2+\fr 14 \left[\beta \cot(\theta'/2) \right]^2 \nonumber \\
&& -  \beta\, \zeta_\perp\,(\mbox{Im}\, a) \,\cot(\theta'/2)
 \sin(\varphi'-\varphi_\zeta).
 \label{crsection-pl}
 \eea
The interference term depends on the transverse polarization of the initial neutron $\bm \zeta_\perp=\zeta_\perp (\cos\varphi_\zeta, \sin\varphi_\zeta,0)$, but not on the longitudinal spin polarization $\zeta_z$ or the helicity $\lambda$. For small scattering angles, $\theta'\to 0$, the second term on the r.h.s. has a singularity of $(1/\theta')^2$, while the third term has a singularity $1/\theta'$.

Due to time-reversal invariance, this single-spin correlation in Eq.(\ref{crsection-pl}) is the same for either initial or final neutron polarization,
and the spin correlation averages to zero after integration with respect to the final neutron's azimuthal angle $\varphi'$.

The standard differential cross section of this process
{\it averaged over spin states of initial neutrons} has a form
 \be
 \fr{d\sigma^{\rm(st)}(\bn,\,\bn',\bm \zeta')}{d\Omega'}=\fr 12 \sum_{\lambda}\left|f_{\lambda \lambda'}(\bn,\,\bn')\right|^2= \fr 12 \left[
 |a|^2+|\bB|^2 + 2  \, (\bB \bm \zeta^{(f)})\,\mbox{Im}\, a \right],
 \label{crsplfinal}
 \ee
where $\bm \zeta'$ is the detected polarization of the final neutron. The polarization
of the final neutron resulting from the scattering process itself~\cite{Schwinger-48} is expressed in terms of strong and electromagnetic amplitudes as
 \be
 \bm \zeta^{(f)} = \fr{2\,\mbox{Im}\, a}{|a|^2+|\bB|^2}\,\bB.
 \ee

\section{3. Scattering of twisted neutrons by a macroscopic target}

\subsection{3.1. Twisted neutrons with a defined $J_z=m$}

Next, we proceed to the case of twisted neutrons and use an approach developed in Ref.~\cite{SIFSS-2015} for the twisted spinor particles. We assume that the incident twisted neutrons propagate along the quantization ($z$) axis and have well--defined values of ({\it i}) a longitudinal linear momentum $p_z$, ({\it ii}) an absolute value of a transverse momentum $|{\bm p}_\perp| \equiv \hbar\varkappa$, and ({\it iii}) a projection of a total angular momentum $J_z = m$, where $m$ is a half--integer. Such \textit{a Bessel state} has, moreover, a definite energy $E = (\hbar^2\varkappa^2 + p_z^2)/(2 m_n)$, with $m_n$ being the neutron mass, and the helicity $\lambda$. The wave function is:
\begin{equation}
      \psi_{\varkappa m p_z \lambda}({\br}) = \int{\frac{{\rm d}^2 {\bp}_\perp}{(2 \pi)^2}} \, a_{\varkappa m}({\bp}_\perp) \,\bii^\lambda
   w^{(\lambda)}(\bn) \, {\rm e}^{\bii {\bp} {\br}/\hbar} \,.
   \label{Besselwf}
\end{equation}
Clearly, the function $\psi_{\varkappa m p_z \lambda}({\br})$ can be considered as a coherent superposition of the plane waves $w^{(\lambda)}(\bn) \, {\rm e}^{\bii {\bp} {\br}/\hbar}$, weighted with the amplitude
\begin{equation}
   \label{eq_a_amplitude}
   a_{\varkappa m}({\bp}_\perp) = \bii^{-m} \, {\rm e}^{\bii m \varphi} \, {\frac{2 \pi}{p_\perp}} \, \delta\left(p_\perp - \hbar \varkappa \right) \, .
\end{equation}
The momenta of these plane--wave components, $${\bp} = \left( {\bp}_\perp, p_z \right) = \left(\hbar  \varkappa \cos\varphi, \hbar \varkappa \sin\varphi, p_z \right),$$ form a surface of a cone with an {\it opening angle}
$ \theta = \arctan (\hbar \varkappa / p_z)$.

Spinor states of the initial and final neutron with helicities $\lambda$ and $\lambda'$ can be expressed as
 \be
 w^{(\lambda)}(\bn)=\sum_{\sigma=\pm 1/2} \bee^{-\bii\sigma \varphi}\,
 d^{\;1/2}_{\sigma \lambda}(\theta)\,  w^{(\sigma)}({\bf e}_z),\;\;
 w^{(\lambda')}(\bn')=\sum_{\sigma'=\pm 1/2} \bee^{-\bii\sigma' \varphi'}\, d^{\;1/2}_{\sigma' \lambda'}(\theta')\,  w^{(\sigma')}({\bf e}_z),
 \label{spinor-w}
 \ee
where $d^{\;1/2}_{\sigma \lambda}(\theta)=\delta_{\sigma \lambda} \cos\left(\theta/2\right) - 2 \sigma \delta_{\sigma, -\lambda} \sin\left(\theta/2\right)$ are the small Wigner $d$-functions and
 \bea
 w^{(1/2)}({\bf e}_z) = \left( \begin{array}{c}
                                   1 \\[0.4cm]
                                   0
                                   \end{array}
                          \right) \, , \: \: \:
   w^{(-1/2)}({\bf e}_z) = \left( \begin{array}{c}
                                   0 \\[0.4cm]
                                   1
                                   \end{array}
                          \right).
 \eea

Using Eq.~\eqref{spinor-w} and the well-known relation
 \be
 \int_0^{2\pi} \fr{d\phi}{2\pi} \, \bee^{\bii (n\phi+z \cos\phi)}= \bii^n\,J_n(z),
 \ee
where $J_n(z)$ is the Bessel function of the first kind, we obtain the evident expressions for the wave-function~\eqref{Besselwf} and the corresponding flux $j_z$ and density $\rho$ of the incoming neutrons (in the cylinder coordinates $r_\perp, \varphi_r,z$):
 \bea
 \label{tw-psi}
  \psi_{\varkappa m p_z \lambda}({\br})&=& \bee^{\bii p_z z/\hbar}\,
  \sum_\sigma \bii^{\lambda -\sigma}
  J_{m-\sigma}(\varkappa r_\perp)\, \bee^{\bii (m-\sigma)\varphi_r}\, d^{\,1/2}_{\sigma \lambda}(\theta)\,w^{(\sigma)}({\bf e}_z),
  \\
  j^{\,(m \lambda)}_z(\br_\perp)&=& \fr{p_z}{m_n}\,\rho^{\,(m \lambda)}(\br_\perp)=
   \fr{p_z}{m_n}\sum_\sigma J^2_{m-\sigma}(\varkappa r_\perp)\,  \left[d^{\,1/2}_{\sigma \lambda}(\theta)\right]^2.
  \label{Flux}
 \eea
Let us consider the limit of these functions at $\theta\to 0$ and the fixed energy $E$ (in this case $\varkappa \to 0$, $p_z\to p=\sqrt{2m_n E}$):
 \be
  \psi_{\varkappa m p_z}(\br)\big|_{\theta \to 0}=
 \delta_{m\lambda}\,w^{(\lambda)}({\bf e}_z) \, {\rm e}^{\bii pz/\hbar},\;\;
 j^{(m,\lambda)}_z(\br_\perp)\big|_{\theta \to 0}=\delta_{m\lambda}\,\fr{p_z}{m_n}.
 \label{plane-limit-wf}
  \ee
In other words, in this limit and at $m=\lambda$ we obtain the standard expressions for the plane-wave neutron flying along $z$~axis with helicity $\lambda$.

Let us consider scattering on a conventional thin-foil target, which we describe as an ensemble of atoms uniformly distributed over the large (compared to the beam's width) transverse extent; we call it {\it a macroscopic target}. If the target is thin, so that one can neglect the neutrons' multiple scattering and attenuation, the scattering cross section can be obtained by the averaging over the atoms' positions in the target w.r.t. the beam axis. Such an averaged cross section represents an incoherent superposition of the standard ones (see Sec. B3 in~\cite{SIFSS-2015}),
 \be
 \fr{d\bar \sigma (\theta, \theta',\varphi', \bm \zeta)}{d\Omega'}=
 \fr{1}{\cos\theta}
 \int_0^{2\pi} \fr{d\sigma^{\rm(st)}(\bn ,\,\bn',\bm \zeta)}{d\Omega'}\;
 \fr{d\varphi}{2\pi}.
 \label{avXsec}
 \ee

To perform the integration in Eq.(\ref{avXsec}), it is useful to expand the vector $\bf B$ in terms of the
unit vectors
 \be
  \bE'_1=(\cos\varphi',\,\sin\varphi',\, 0),\ \bE'_2=(-\sin\varphi',\,\cos\varphi',\, 0),\;\;
\bE'_3=(0,\,0,\, 1),\;\; \bE'_i \bE'_k=\delta_{ik}
  \ee
as follows
 \be
 \bB=\fr{\beta}{2(1- \bn \bn')} \left\{
 (sc'\bE'_1- cs'\bE'_3)\,\sin(\varphi-\varphi') + [cs'-sc'\cos(\varphi-\varphi')]\bE'_2\right\},
 \label{B-evid}
   \ee
where $s\equiv\sin\theta$, $c\equiv\cos\theta$, $s'\equiv\sin\theta'$, $c'\equiv\cos\theta'$,
and to use the relation
 \be
 \bB^2=\beta^2 \left[ \fr{1}{2(1- \bn \bn')}- \fr 14\right]
 \ee
with
 \be
 (\bn-\bn')^2=2(1- \bn \bn')=2(1-cc' - ss'\cos(\varphi-\varphi')).
 \ee
Using these relations, we obtain
 \bea
 && \int_0^{2\pi}
 \left(\fr{{\bf B}(\varphi)}{\beta} \right)^2 \, \fr{d\varphi}{2\pi}=
 \fr{1}{2 |\cos\theta-\cos\theta'|} - \fr 14
 =G(\theta, \theta'),
 \nn
 \\
 &&\int_0^{2\pi}\,
 \fr{{\bf B}(\varphi)}{\beta}\, \fr{d\varphi}{2\pi}= \fr 12 \,g(\theta, \theta')\, \bE'_2,\;\;
 g(\theta, \theta')
 =\left\{\begin{array}{c}
    \cot(\theta'/2)\;\;\;\;\;\mbox{at}\;\;\theta'>\theta  \\
    -\tan(\theta'/2)\;\;\mbox{at}\;\;\theta'<\theta
    \nn
  \end{array}\right. .
  \label{funcGg}
  \eea
Note that the function $G(\theta, \theta')$ is singular:
 \be
 G(\theta, \theta')\to \fr{1}{2|\theta'-\theta|\,\sin\theta}\;\;\mbox{at}\;\;\theta'\to \theta.
 \ee

As a result, we get ($c.f.$ Eq.(\ref{crsection-pl}))
 \be
\label{cr-section-averaged-R}
 \fr{d\bar \sigma (\theta, \theta',\varphi', \bm \zeta)}{d\Omega'} =
   \fr{|a|^2}{\cos\theta}\left[1  +R_{\rm em} -\,\fr{\mbox{Im}\, a}{|a|}\, R_{\rm int}\,\zeta_\perp
    \,\sin(\varphi'-\varphi_\zeta)\right],
      \ee
where
\be
 R_{\rm em}=\varepsilon^2 G(\theta, \theta'),\;\;R_{\rm int}=\varepsilon g(\theta, \theta')
 \label{Ratio}
 \ee
This cross section is still independent of $\zeta_z$ and it coincides with Eq.\eqref{crsection-pl} in the
standard limit $\theta\to 0$. Such a behavior is expected since in this limit $G(\theta, \theta')\to (1/4) \cot^2(\theta'/2)$ and $g(\theta, \theta')\to  \cot(\theta'/2)$.

\begin{figure}[t]
	\center
	\includegraphics[width=0.5\linewidth]{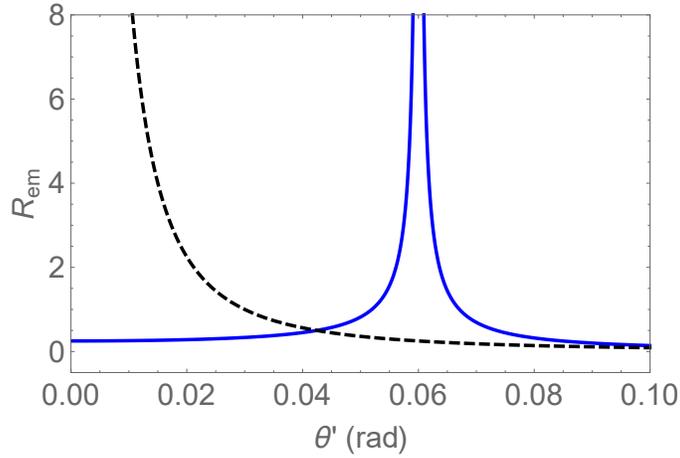}
	\caption{The functions $R_{\rm em}$ (blue solid line) from
Eq.~\eqref{Ratio} and its plane-wave limit (black dashed line) plotted vs. the neutron scattering angle $\theta'$ for the opening angle $\theta=0.06$ rad and parameter $\varepsilon=0.03$.}
\label{Fig1}
\end{figure}

In Fig.~\ref{Fig1} we present the function $ R_{\rm em}(\theta') $ which corresponds to a relative contribution of the electromagnetic interaction. Unlike the Schwinger cross section \eqref{crsection-pl}, this function has an angular singularity of $1/|\theta'-\theta|$ at $\theta' \to \theta$. This shift to the non-vanishing scattering angles is potentially useful for experimental analysis of the small-angle scattering. Indeed, thanks to this property, the singular region is shifted from the small angles $\theta'\to 0$, which may be difficult to access experimentally, to the larger values, $\theta'\to \theta$, which can be controlled by the opening angle $\theta$ of the incoming twisted neutrons. Practically, this method would depend on experiment's ability to reach sufficiently large values of the opening angle  $\theta$.

\begin{figure}[t]
	\center
	\includegraphics[width=0.5\linewidth]{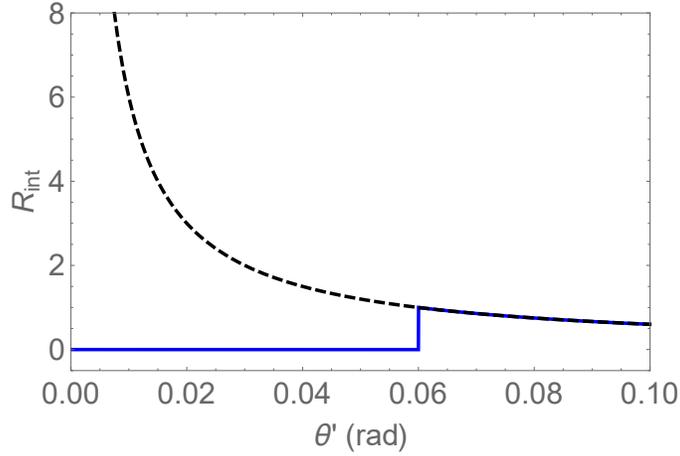}
	\caption{The functions $R_{\rm int}$ (blue solid line) from
Eq.~\eqref{Ratio} and its plane-wave limit (black dashed line) plotted vs. the neutron scattering angle $\theta'$ for $\theta=0.06$ rad and $\varepsilon=0.03$.}
\label{Fig2}
\end{figure}

In Fig.~\ref{Fig2} we present the function $R_{\rm int}(\theta')$ which describes interference of the electromagnetic amplitude and the nuclear one as well as the corresponding function for the standard case.
In the region $\theta'\ge\theta$, the function $R_{\rm int}(\theta')$ coincides with its standard limit, but
 this function experiences a step-like drop for the angles $\theta'\le\theta$ that potentially can be observed in experiments.

\subsection{3.2. Twisted neutrons in a superposition of two vortex states}

Let us take now a coherent superposition of two Bessel states with the different projections $m_1$ and $m_2$, but with the same helicity $\lambda$ and the same values of $p_z$ and $\varkappa$.
Such a superposition can be generated experimentally \cite{exp-2015, exp-2018}, and it is described by the following wave function:
\bea
   \label{eq_wave_function_twisted_electrons_superposition}
   \psi^{\rm (2 \, tw)}({\br}) &=& c_1 \psi_{\varkappa m_1 p_z \lambda}({\br}) + c_2 \psi_{\varkappa m_2 p_z \lambda}({\br}),\;\; \nonumber \\
   c_n &=& |c_n| {\rm e}^{i \alpha_n} \, , \: \: \: |c_1|^2 + |c_2|^2 = 1 \, .
\eea

With the help of this expression, we find the averaged differential cross section in the form
 \be
 \fr{d\bar \sigma (\theta, \theta',\varphi', \bm \zeta)}{d\Omega'}=
 \fr{1}{\cos\theta}
 \int_0^{2\pi} \fr{d\sigma^{\rm(pl)}(\bn ,\,\bn',\bm \zeta)}{d\Omega'}\,
 \Phi(\varphi, \Delta m, \Delta \alpha)\,
 \fr{d\varphi}{2\pi},
 \label{2-cr-section-averaged}
 \ee
where the function $\Phi$ is defined as
\begin{equation}
   \label{eq_G_function}
    \Phi(\varphi, \Delta m, \Delta \alpha) = 1 + 2|c_1 c_2| \, \cos \left[(\varphi - \pi/2)\, \Delta m +\Delta \alpha) \right] \, .
\end{equation}

As a result, we obtain
  \bea
  \label{cs-two-projection}
 &&\fr{d\bar \sigma (\theta, \theta',\varphi', \bm \zeta)}{d\Omega'} = \nn \\
&& \fr{1}{\cos\theta}\left\{ \mathcal{A}+|c_1c_2|\left(\beta^2\,\mathcal{B} +
 2(\mbox{Im}\, a)\,\beta\, (\bm\zeta{\bf C)} \right)\right\},
  \nn\\
  &&\mathcal{A} = |a|^2+\beta^2 G(\theta, \theta')+(\mbox{Im}\, a)\,\beta\, (\bm\zeta\bE_2)\,
  g(\theta, \theta'),
  \nn
  \\
  &&\mathcal{B} = \fr{\cos\gamma}{|c-c'|}\; [T(\theta, \theta')]^{|\Delta m|},
  \,
  \\
  && {\bf C} = \left[ \fr{\Delta m}{|\Delta m|}\,\left( - \fr{c'}{s'}\,\bE'_1+ \bE'_3\right)\,\sin\gamma
  + \fr{c-c'}{|c-c'|}\, \bE'_2 \,\cos\gamma \right] \nn \\
  &&\times [T(\theta, \theta')]^{|\Delta m|},
  \nn
    \eea
where
 \bea
 &&\gamma =(\varphi' - \pi/2)\, \Delta m +\Delta \alpha,\;\; \nn \\
 &&T(\theta, \theta')= \left(\fr{\tan(\theta/2)}{\tan(\theta'/2)}\right)^{\pm 1}
 \;\; \mbox{for}\;\; \theta' \gtrless \theta.
  \nn
 \eea
In contrast to Eq.~(\ref{cr-section-averaged-R}), derived for a single-$m$ incident beam, this cross section depends on the differences of the total angular momenta, $\Delta m = m_2 - m_1\neq 0$, and of the states' phases, $\Delta \alpha = \alpha_2 - \alpha_1$. This $\Delta m$ and $\Delta \alpha$ dependence translates directly into the angular- and polarization properties of the scattered neutrons.
In particular:

({\it i}) The cross section (\ref{cs-two-projection}) depends not only on the neutron's transverse polarization $\bm \zeta_\perp$, but also on the longitudinal one $\zeta_z$.
It leads to the following {\it longitudinal spin asymmetry}:
 \bea
 A_{\zeta_z}&=&\fr{d\bar \sigma (\zeta_z=+1)-d\bar \sigma (\zeta_z=-1)}
 {d\bar \sigma (\zeta_z=+1)+d\bar \sigma (\zeta_z=-1)}\nn \\ &=&
  \fr{2|c_1c_2|\,(\mbox{Im} a)\,\beta\,({\bf C} \bE'_3) }{|a|^2+\beta^2 \left[
  G(\theta, \theta')+|c_1c_2|\,\mathcal{B}\right]}.
  \label{A-longitudinal}
 \eea

\begin{figure}[h!]
\centering
\includegraphics[width=0.5\linewidth]{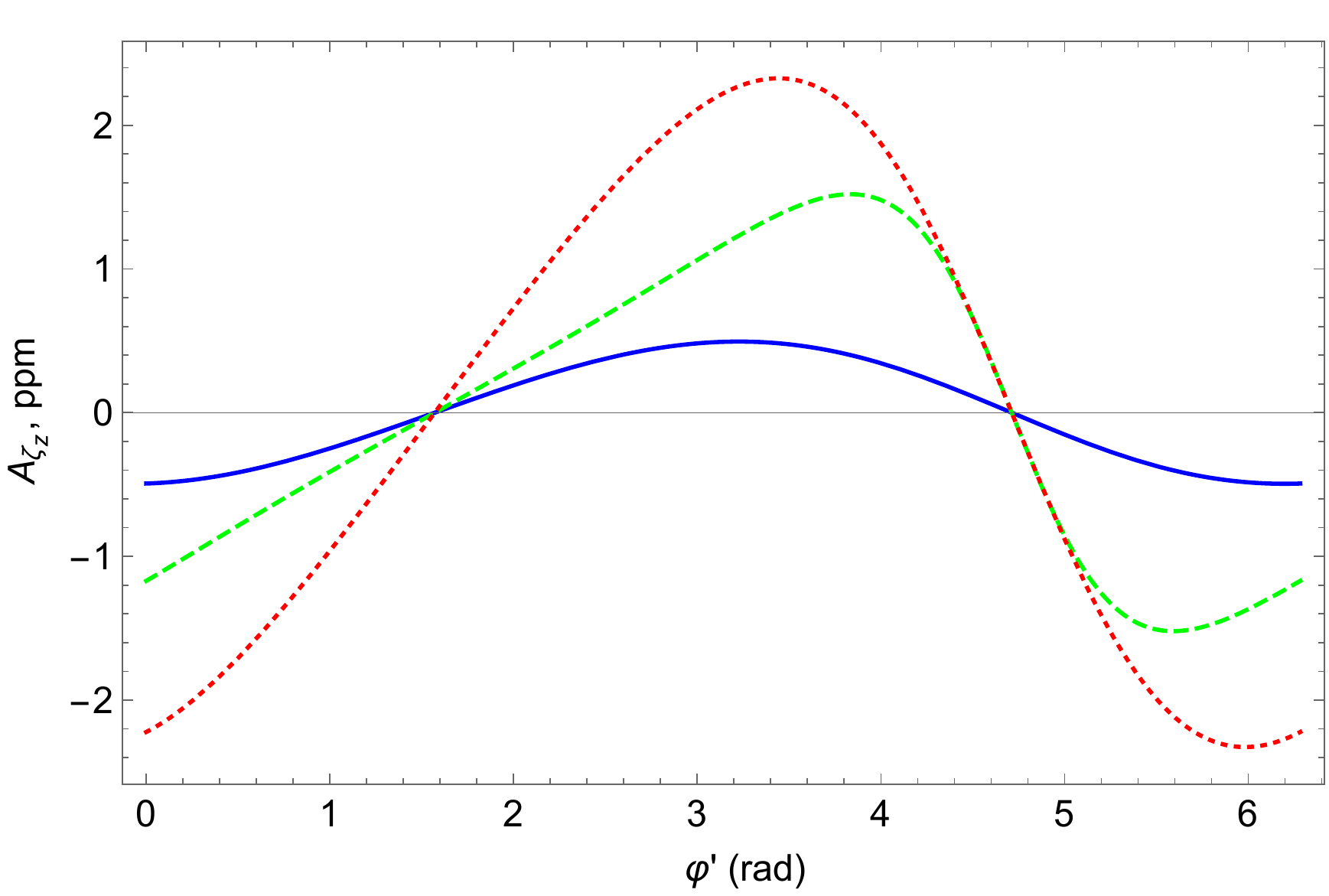}
\caption{A longitudinal spin asymmetry~\eqref{A-longitudinal} plotted vs. the neutron scattering
azimuthal angle $\varphi'$ for $\varepsilon=0.03$, $2 c_1c_2=1$, $\Delta m=1$, and for $\theta'=0.005$ rad (blue solid line), $\theta'=0.025$ rad (green dashed line), $\theta'=0.045$ rad (red dotted line).}
\label{Fig3}
\end{figure}

\begin{figure}[h!]
\centering
\includegraphics[width=0.5\linewidth]{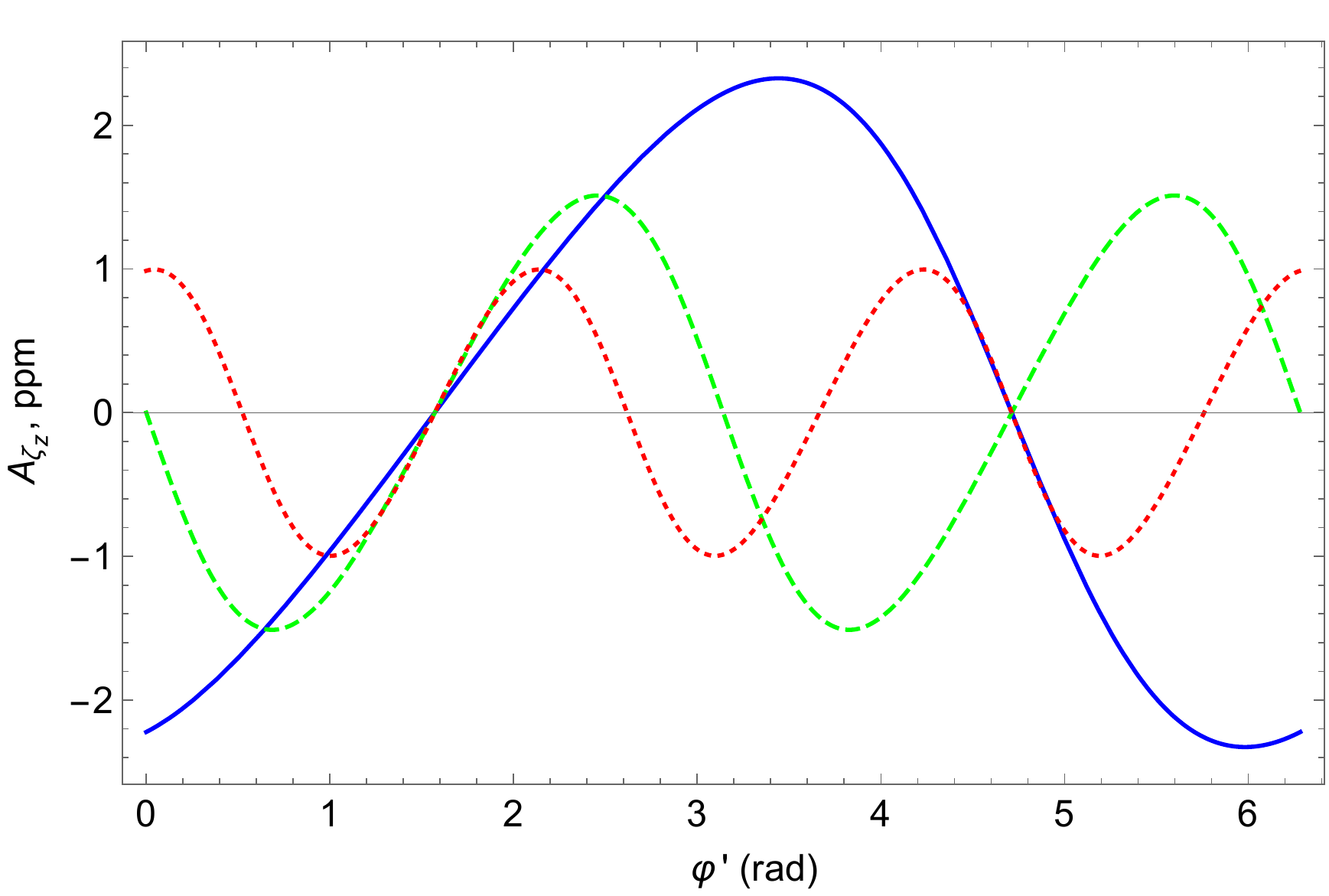}
\caption{The longitudinal spin asymmetry~\eqref{A-longitudinal} plotted vs. the neutron scattering
azimuthal angle $\varphi'$ for $\varepsilon=0.03$, $2 c_1c_2=1$, $\theta'=0.045$, and for $\Delta m=1$ (blue solid line), $\Delta m=2$ (green dashed line), $\Delta m=3$ (red dotted line).}
\label{Fig4}
\end{figure}

Figures~\ref{Fig3} and \ref{Fig4} show this asymmetry for different values of the parameters. For thermal neutrons and a gold target (see Eq.~\eqref{parameters-for-Au}), the predicted asymmetry amounts to {\it a few ppm}, which is in a range currently accessible for experiments on the hadronic parity violation \cite{Musolf-review}, for which the above asymmetry may be a source of unwanted systematics, provided that the neutron beam becomes twisted due to uncontrolled interactions. However, as we show below, averaging over the azimuthal scattering angle $\varphi'$ eliminates the dependence on $\zeta_z$, which provides an approach to correct for this kind of systematics. Azimuthal angular coverage for the neutron-scattering detectors would be essential to deal with this systematic effect in parity-violation measurements.

\begin{figure}[h!]
\centering
\includegraphics[width=0.5\linewidth]{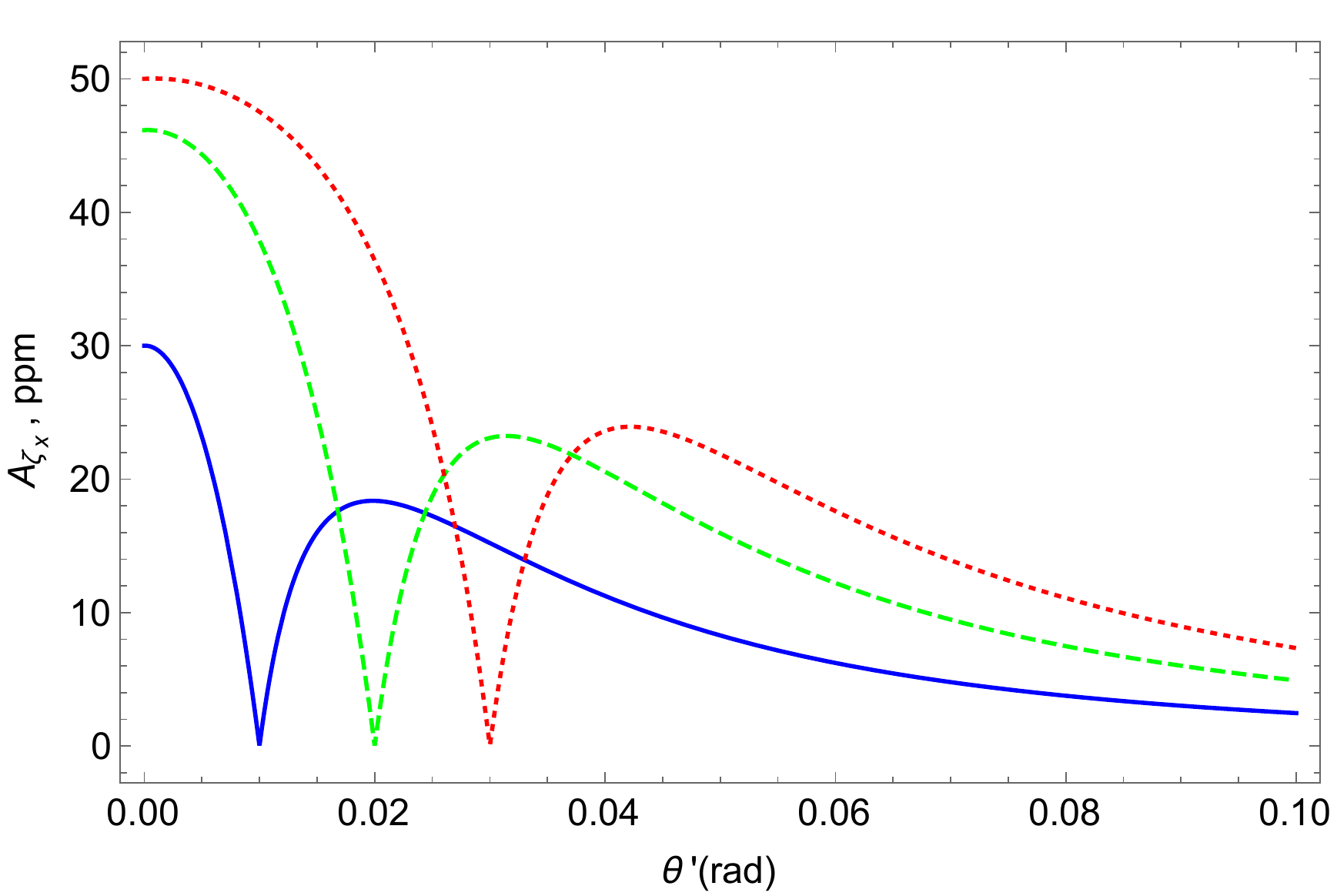}
\caption{A transverse spin asymmetry plotted vs. the neutron scattering
angle $\theta'$ for $\varepsilon=0.03$, $2 c_1c_2=1$, $\Delta m=1$, and for $\theta=0.01$ (blue solid line),
$\theta=0.02$ (green dashed line), $\theta=0.03$ (red dotted line).}
\label{Fig5}
\end{figure}

Analogously to Eq.~\eqref{A-longitudinal}, one can define a quantity $ A_{\zeta_x}$, which we call the {\it transverse spin asymmetry}.  We show this asymmetry in Fig.~\ref{Fig5} for a sample set of parameters.

({\it ii})
Let us discuss the properties of the differential cross section~\eqref{cs-two-projection} averaged over the azimuthal angle $\varphi'$ of the final neutron. For this aim, we introduce the following notation:
 \be
 \langle F \rangle = \int_0^{2\pi} F\; \fr{d\varphi'}{2\pi}.
 \label{averaging}
 \ee
The averaged cross section reads
 \bea
  \label{cs-two-projection-averaged}
 \left\langle
 \fr{d\bar \sigma (\theta, \theta',\varphi', \bm \zeta)}{d\Omega'}
 \right\rangle &=&
 \fr{1}{\cos\theta}\Big (|a|^2+\beta^2 G(\theta, \theta')+ \nn \\ &&
 2|c_1c_2|\, (\mbox{Im}\, a)\,\beta\, \bm\zeta \langle{\bf C}\rangle \Big),
  \eea
where
 \be
 \label{Caveraged}
  \langle{\bf C}\rangle =
     \frac 12 \, \left(\fr{c'}{s'} - \fr{c-c'}{|c-c'|} \right)
 T(\theta, \theta')\, (\cos\Delta\alpha,\, \mp \sin\Delta\alpha,\,0)
  \ee
for $\Delta m = \pm 1$, and $\langle{\bf C}\rangle =0$ otherwise. The fact that this  spin observable is nonzero can be understood as due to an effect of superposition between two vortex states that define a new plane with an orientation fixed by a phase difference $\Delta\alpha$. Then the transverse spin with respect to this plane contributes to the scattering asymmetry, while the neutron scattering plane -- the only plane available for non-twisted neutrons -- becomes redundant.
It is seen that this cross section depends on $\bm \zeta_\perp$ and on $\mbox{Im}\, a$ at $\Delta m =\pm 1$, but it is independent of the longitudinal polarization.

\begin{figure}[h!]
\centering
\includegraphics[width=0.4\linewidth]{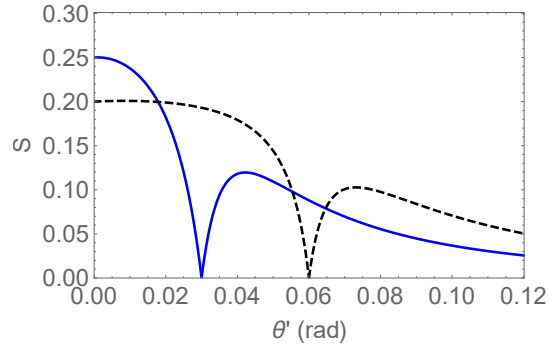}
\caption{The function $S$ (defined in Eq.~\eqref{S-polarization-final}) plotted vs. the neutron scattering angle $\theta'$ for $\varepsilon=0.03$, $2 |c_1c_2|=1$, $\Delta m=1$, and for  $\theta=0.03$ rad (blue solid line),
$\theta=0.06$ rad (black dashed line).}
\label{Fig6}
\end{figure}

If the initial neutron is unpolarized,
then its polarization after the scattering is
 \be
 \langle\bm\zeta^{(f)} \rangle = -\fr{\mbox{Im}\, a}{|a|}\, S \, (\cos\Delta\alpha,\, \mp \sin\Delta\alpha,\,0),
 \ee
where
 \be
 S=\fr{|c_1c_2| \varepsilon}{1+\varepsilon^2 G(\theta, \theta')}
 \left(\fr{c'}{s'} - \fr{c-c'}{|c-c'|} \right)
 T(\theta, \theta')
 \label{S-polarization-final}
 \ee
for $\Delta m = \pm 1$, and $S =0$ otherwise. In Fig.~\ref{Fig6} one can see that $|\langle\bm\zeta^{(f)} \rangle| \sim 0.1 \; \fr{|\mbox{Im}\, a|}{|a|}$ for $\theta \sim \varepsilon$,  i.\,e. the predicted effect is of the order of {\it tens of ppm} for the thermal neutrons and the gold target (see Eq.(\ref{parameters-for-Au})).

\section{4. Scattering of twisted neutrons by a single nucleus}

Let the single-$m$ neutrons be scattered by a nucleus located in the transverse ($xy$) plane at a definite impact parameter ${\bb}= (b_x, b_y ,0)=b\,(\cos\varphi_b,\,\sin\varphi_b,\,0)$.
Using the neutron's wave function (\ref{Besselwf}), and Eq.(\ref{1}), we find the following scattering amplitude:
 \bea
   \label{eq_scattering_amplitude_twisted_wave}
   F_{\lambda \lambda'}^{(m)}(\theta, \theta',\varphi', {\bb})&&  = \bii^{\lambda-m}
   {\rm e}^{- \bii {\bp}'_\perp {\bb}/\hbar }\, \nn \\
   &&\times\int_{0}^{2\pi} \frac{{\rm d}\varphi}{2 \pi} \,
   {\rm e}^{\bii m \varphi + \bii {\bp}_\perp {\bb}/\hbar }
   f_{\lambda \lambda'}(\bn,\,\bn'),
 \eea
where the factor ${\rm exp}(\bii {\bp}_{\perp} {\bb}/\hbar )$ specifies the lateral position of the nucleus with respect to the beam.

 Using Eqs.~\eqref{spinor-w}, \eqref{B-evid}  and introducing quantities
 \bea
 \label{Asigma}
 A^{(\sigma)}(m,\varkappa, \bb)&=& \int_0^{2\pi} \fr{d\varphi}{2\pi}\,
 \bee^{\bii [(m-\sigma)\varphi+\varkappa b \,\cos(\varphi-\varphi_b)]}\, a=
 a\,  \bee^{\bii (m-\sigma)(\varphi_b+\pi/2)} \, J_{m-\sigma}(\varkappa b),
 \\
 \bB^{(\sigma)}(m,\varkappa, \bb)&=& \int_0^{2\pi} \fr{d\varphi}{2\pi}\,
 \bee^{\bii [(m-\sigma)\varphi+\varkappa b \,\cos(\varphi-\varphi_b)]}\, \bB(\varphi),
   \label{Bsigma}
   \eea
we rewrite the above equation in the form
 \be
 F_{\lambda \lambda'}^{(m)}(\theta, \theta',\varphi', {\bb})  = \bii^{\lambda-m} {\rm e}^{- \bii {\bp}'_\perp {\bb}}\,
 \sum_{\sigma=\pm 1/2}\,d^{\;1/2}_{\sigma \lambda}(\theta)\, w^{(\lambda')\dag}(\bn') \left[ A^{(\sigma)}+\bii \bB^{(\sigma)} \bm\sigma \right]
 w^{(\sigma)}({\bf e}_z).
  \ee
This amplitude coincides (up to the inessential factor ${\rm e}^{- \bii {\bp}'_\perp {\bb}}$) with the standard one~\eqref{1} in the limit $\theta\to 0$ since in this limit $\bn\to {\bf e}_z$, $d^{\;1/2}_{\sigma \lambda}(\theta)\to \delta_{\sigma \lambda}$, $A^{(\sigma)}\to a\,\delta_{m\sigma}$ and
$\bB^{(\sigma)}\to \bB\,\delta_{m\sigma}$.

The angular distributions of the scattered neutrons can be obtained by squaring this amplitude.
Such a distribution {\it summed} over helicities of final neutrons
 \be
 \label{AngularDistribution-1}
 W^{(m)}_{\lambda}(\theta, \theta', \varphi', \bb)=\sum_{\lambda'}
 \left|F_{\lambda \lambda'}^{(m)}(\theta, \theta',\varphi', {\bb}) \right|^2
 \ee
is considered in detail in the Appendix A and the specific case $\bb=0$ in the Appendix B. Here we only discuss this distribution {\it averaged} over the azimuthal angle of the final neutrons using the notation~\eqref{averaging}:

 \begin{widetext}
\bea
 \label{AngularDistribution-2}
& W^{(m)}_{\lambda}(\theta, \theta',  \bb)= \sum\limits_{\lambda'} \left\langle
 \left|F_{\lambda \lambda'}^{(m)}(\theta, \theta',\varphi', {\bb}) \right|^2\right\rangle
 = \fr 12\, \Sigma^{(m)} + \lambda\, \Delta^{(m)},
  \\
& \Sigma^{(m)}=|a|^2 \left(J^2_{m-1/2}(\varkappa b)+ J^2_{m+1/2}(\varkappa b) \right)
  + \sum\limits_{\sigma} \left\langle (\bB^{(\sigma)*}\bB^{(\sigma)})
  -2\sigma\mbox{Im} \left(\bB^{(\sigma)*}\times \bB^{(\sigma)}\right)_z \right\rangle,
 \cr
& \Delta^{(m)}=
\left(|a|^2\cos\theta- (\mbox{Re}\, a) \,\beta\, h(\theta, \theta') \right)\,\left(J^2_{m-1/2}(\varkappa b)- J^2_{m+1/2}(\varkappa b) \right) +
 \cr
 & +  \cos\theta \sum\limits_{\sigma} \left\langle 2\sigma (\bB^{(\sigma)*}\bB^{(\sigma)})
  -\mbox{Im} \left( \bB^{(\sigma)*}\times \bB^{(\sigma)}\right)_z \right\rangle -
 \sin\theta\,\left\langle \mbox{Im} \left( \bB^{(1/2)*}\times \bB^{(-1/2)}\right)_x  -
  \mbox{Re} \left( \bB^{(1/2)*}\times \bB^{(-1/2)}\right)_y \right\rangle,
  \nn
 \eea
\end{widetext}

where
 \be
h(\theta, \theta')=g(\theta', \theta)\,\sin\theta= \left\{\begin{array}{c}
    -2\sin^2(\theta/2)\;\;\mbox{at}\;\;\theta'>\theta  \\
    \;\, 2\cos^2 (\theta/2)\;\;\mbox{at}\;\;\theta'<\theta
    \end{array} \right\}
    =\mp (1\mp \cos\theta) \;\;\mbox{for}\;\;  \theta'\gtrless \theta .
 \ee

As a result, we obtain a nonvanishing {\it helicity asymmetry},
 \be
 \label{Alambda}
 A_\lambda = \fr{ W_{\lambda=1/2}^{(m)} - W_{\lambda=-1/2}^{(m)} }
 { W_{\lambda=1/2}^{(m)}  + W_{\lambda=-1/2}^{(m)} } =
 \fr{\Delta^{(\rm m)}} {\Sigma^{(\rm m)}}.
 \ee

\begin{figure}[h!]
\center
\includegraphics[width=0.99\linewidth]{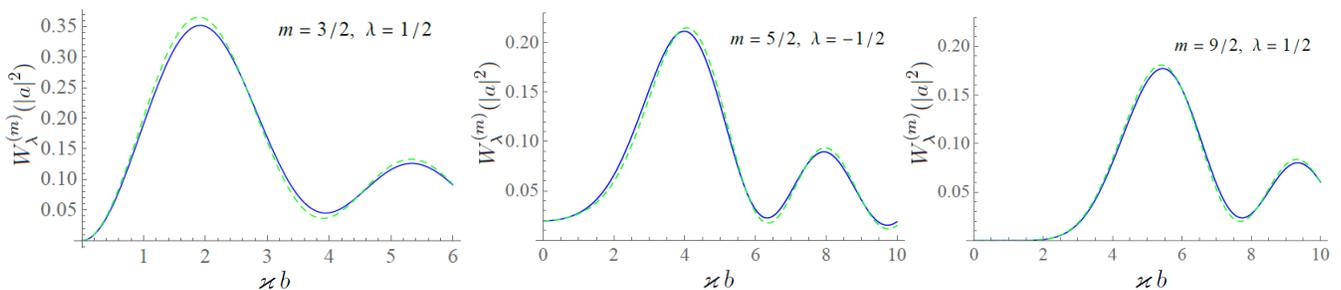}%
\caption{The distribution (\ref{AngularDistribution-2}) in units $|a|^2$ as a function of the $^{197}_{79}\rm Au$ nucleus position $b$ for $\theta^{\prime} = 0.04$~rad, $\theta=0.07$~rad, and $\varepsilon=0.03$. The case ${\rm Im}\;a > 0$ is shown by blue solid lines, while the case ${\rm Im}\;a < 0$ is shown by green dashed lines.}
\label{Fig7}
\end{figure}

In contrast to Eq.\eqref{crsection-pl}, the interference term in (\ref{AngularDistribution-2}) depends on the initial neutron's {\it helicity} and on the {\it real part} of the nuclear amplitude (and, therefore, on its phase $\text{Arg}\, a$), even after the azimuthal averaging -- see \eqref{AngularDistribution-2}.
The angular distributions (\ref{AngularDistribution-2}) are plotted in Fig.~\ref{Fig7} for an $^{197}_{79}\rm Au$ nucleus as a function of its position $b$. The scattering angle is chosen as $\theta^{\prime} = 0.03$~rad for which the electromagnetic and strong amplitudes equally contribute to the cross section for the plane-wave neutrons. One can see that the former contribution dominates in the beam center ($b\to$0), where the interference between two amplitudes is most pronounced.

The asymmetry $A_\lambda$ is a periodic function of $b$ and of the amplitude's phase $\text{Arg}\, a$;
it can reach {\it tens of percent} for a wide range of parameters, as shown in Ref.~\cite{Afanasev-PRC19}.
Note that outside the cone opening angle, at $\theta^{\prime} > \theta$, the sensitivity to the phase practically vanishes,
so that in order to probe both the real and the imaginary parts of the amplitude one needs to perform measurements at the small angles $\theta^{\prime} < \theta$, which is feasible.

\section{5. Scattering of twisted neutrons by a mesoscopic nuclear target}

Until now we have discussed the scattering for two extreme cases
of either a single-nucleus or a macroscopic (infinitely wide)
target. In a more realistic experimental scenario, a neutron beam
collides with a well--localized mesoscopic atomic target. In order
to account for geometrical effects in such a scenario, we describe
a target as an incoherent ensemble of potential centers. The
density of the scatterers in the transverse ($xy$) plane is
characterized by a distribution function $n({\bf b})$, which is
normalized as follows:
\begin{equation}
\label{eq:electron_distribution}
\int n({\bf b}) \, {\rm d}^2b = 1 \, .
\end{equation}
For the numerical analysis below we take $n({\bf b})$ to be a Gaussian function:
\begin{equation}
   \label{eq:Gaussian_distribution_2}
   n({\bf b}-{\bf b}_t) = \frac{1}{2 \pi \sigma_t^2} \, {\rm e}^{-(\bb - \bb_t)^2/(2 \sigma_t^2)}
   \, .
\end{equation}
This distribution is sharply peaked at the impact parameter $\bb={\bb}_t$ (the centre of the target) if its dispersion $\sigma_t$ is small.

\begin{figure}[h!]
\center
\includegraphics[width=0.5\linewidth]{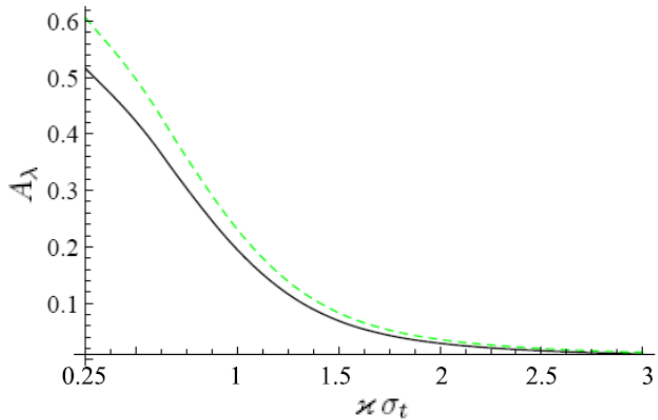}%
\caption{The helicity asymmetry as a function of $\varkappa \sigma_t$ where $\sigma_t$ is a width of the $^{197}_{79}\rm Au$ mesoscopic target for $m=1/2$, $\theta^{\prime} = 0.03$~rad, $\theta=0.06$~rad, and $\varepsilon=0.03$, $b_t=\varphi_t=0$. The case ${\rm Im}\;a > 0$ is shown by the black solid line, while the case ${\rm Im}\;a < 0$ is shown by the green dashed line.}
\label{Fig8}
\end{figure}

\begin{figure}[h!]
\center
\includegraphics[width=0.5\linewidth]{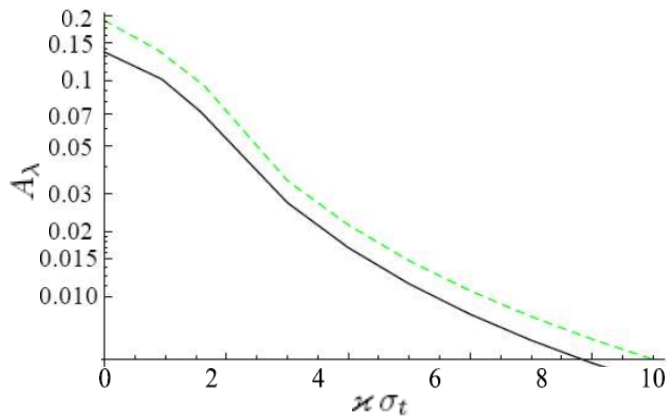}%
\caption{The helicity asymmetry as a function of $\varkappa \sigma_t$ where $\sigma_t$ is a width of the $^{197}_{79}\rm Au$ mesoscopic target for $m=5/2$, $\theta^{\prime} = 0.04$~rad, $\theta=0.07$~rad, and $\varepsilon=0.03$, $b_t=\varphi_t=0$. The case ${\rm Im}\;a > 0$ is shown by the black solid line, while the case ${\rm Im}\;a < 0$ is shown by the green dashed line.}
\label{Fig9}
\end{figure}

There are two limiting cases: {\it(i)} when the target is much wider than the incident beam, $\sigma_t \gg 1/\varkappa$, and {\it(ii)} when it is narrower, $\sigma_t \lesssim 1/\varkappa$.
When averaging over the impact parameters ${\bf b}$, the above features of the single-nucleus regime survive when the target is sub-wavelength sized or even if its width does not exceed that of the beam,
whereas for the macroscopic target the above spin asymmetries vanish, see Eq.~(\ref{cr-section-averaged-R}). For an intermediate case of a mesoscopic target, the spin asymmetry survives
but its values decrease as $\sigma_t\varkappa$ grows. In order to give quantitative estimates, we take a realistic example of a Gaussian target with $\sigma_t \sim (1/\varkappa) - (10/\varkappa)$
and the angles $\theta^{\prime} < \theta\sim 1^\circ - 10^\circ$. The helicity asymmetry reaches the values of
\be
\label{Al}
|A_\lambda| \approx 10^{-3}-10^{-1}
\ee
for a wide range of parameters, as we show in Fig.\ref{Fig8} and Fig.\ref{Fig9}. Note that the asymmetries even some 2 or 3 orders of magnitude
smaller can in principle be measured, as the current experiments aim at much lower values than (\ref{Al}), down to $10^{-8}$ \cite{Gericke08,Musolf-review}.

Thus, scattering off the well-localized targets -- say, of $\sigma_t \gtrsim 10\,\text{nm}-1\,\mu\text{m}$ in width for the neutron wave packets
with the wavelength of $0.1 - 100$ nm and the transverse coherence length of $1/\varkappa \gtrsim 1\,\text{nm}-10\,\mu\text{m}$ \cite{exp-2018}
-- reveals dependence on the neutron's helicity and allows one to probe the nuclear amplitude's real part already in the Born approximation,
whereas with a single beam of the delocalized plane-wave neutrons such a dependence arises beyond the tree level only.
This method for high-precision measurements of the complex amplitude for non-vanishing scattering angles is alternative and complementary
to the neutron interferometry and to the neutron gravity reflectometry.

Conventional Schwinger asymmetry can be enhanced for thermal neutrons due to presence of the nuclear resonances, as was shown experimentally \cite{Felcher1975}.
Extension of our formalism to the nuclear resonance region is straightforward, in which case we have to use an appropriate parameterization for the nuclear amplitude $a$,
and use a partial-wave expansion and angular integration in Eq.(33) in order to account for the non $S$-wave resonances.

\section{6. Conclusion}

We have presented a theoretical formalism for elastic scattering of the twisted neutrons by a nucleus and nuclear targets and predict new effects for the cross section and spin asymmetries. Our approach is based on expansion of the twisted beam's wave function in terms of the plane waves and it leads to representation of the twisted-scattering amplitude in a form of a superposition of the plane-wave amplitudes. The results are presented for the kinematics of the Schwinger scattering \cite{Schwinger-48} characterized by a spin asymmetry due to interference between the strong and electromagnetic scattering amplitudes.

The following observable effects are predicted that are unique for the twisted-neutron scattering on nuclei:

(a) For the macroscopic targets, the scattering cross section has a different angular dependence, which is peaked at non-zero scattering angles, as opposed to the non-twisted case;

(b) For the macroscopic targets and a beam that is a superposition of angular momentum states differing by one unit of $\hbar$, the cross section develops a spin asymmetry that depends on the azimuthal scattering angle and a longitudinal component of neutron's spin. This observable is forbidden for non-twisted neutrons by parity conservation;

(c) For scattering on a single nucleus, provided that the target's location is resolved with respect to the twisted neutron's wavefront, the scattering spin asymmetry is due to both the longitudinal and transverse spin
and, in addition, the spin asymmetries have contributions from both the real and absorptive parts of the nuclear amplitude. These features of the spin asymmetries still hold for the realistic mesoscopic targets.

The predicted spin asymmetries range from $10^{-6}$ to $10^{-1}$ for relevant parameters and are detectable in experimental conditions similar to those used for parity-violating measurements \cite{Gericke08,Musolf-review}. Whereas generation of the twisted neutrons was experimentally demonstrated for lower fluxes \cite{exp-2015,exp-2018,exp-2019}, it will be desirable to use the high-flux sources of twisted neutrons in order to achieve sufficient statistical accuracy of the future measurements.

In summary, we have demonstrated that twisted neutrons can be used as a new tool to probe nuclear scattering amplitudes at low energies and to access observables that are otherwise forbidden by the symmetry for non-twisted beams.

{\it Acknowledgements} -- We would like to thank D.~Pushin, W.\,M.~Snow, and A.~Surzhykov for useful discussions.The work of D.V.K. and V.G.S. is supported by the Russian Science Foundation (Project No.\,17-72-20013).

\section{Appendix A. Angular distributions of the scattered neutrons
 }

The distribution defined in Eq.(\ref{AngularDistribution-1})  reads
 \bea
 \label{A-AngularDistribution-1}
 W^{(m)}_{\lambda}(\theta, \theta', \varphi', \bb)&=&\sum_{\lambda'}
 \left|F_{\lambda \lambda'}^{(m)}(\theta, \theta',\varphi', {\bb}) \right|^2
  \\
 &=& \fr 12 \left[D^{(1/2)}+D^{(-1/2)} \right] + \lambda \left[D^{(1/2)}-D^{(-1/2)} \right]\, \cos\theta
   \nn
   \\
  &-& 2\lambda\,\left[\mbox{Im} C_x^{(1/2,- 1/2)} -\mbox{Re} C_y^{(1/2, -1/2)}\right] \,\sin\theta.
   \nn
 \eea
Here we use the quantities
 \bea
   {\bf C}^{(\tilde\sigma\sigma)}&=& \bB^{(\tilde\sigma)*}\times \bB^{(\sigma)}+
   {\bf I}^{(\tilde\sigma\sigma)},\;\; {\bf I}^{(\tilde\sigma\sigma)}=
   A^{(\tilde\sigma)*}\,  \bB^{(\sigma)}-\bB^{(\tilde\sigma)*}\,A^{(\sigma)}
    \\
   {D}^{(\sigma)}&=&
     \left|A^{(\sigma)}\right|^2 + \bB^{(\sigma)*}\bB^{(\sigma)}-2\sigma \mbox{Im} C_z^{(\sigma \sigma)}
    \eea
with the properties
 \bea
 C_z^{(\sigma \sigma)}=\bii\,\mbox{Im}\, C_z^{(\sigma \sigma)},\;\; {\bf C}_\perp^{(\sigma,- \sigma)}=
 -\left({\bf C}_\perp^{(-\sigma, \sigma)} \right)^*.
 \eea

 Let us define contributions from the nuclear and electromagnetic interactions and their interference as
 $$
 W^{(m)}_{\lambda}(\theta, \theta', \varphi', \bb)=W^{(m,\rm nucl)}_{\lambda}+
 W_{\lambda}^{(m,\rm em)}+W_{\lambda}^{(m,\rm int)},
 $$
where
 \bea
 W^{(m,\rm nucl)}_{\lambda}&=&\sum\limits_{\sigma} \left(\frac{1}{2} + 2\sigma\,\lambda\cos\theta\right) |A^{(\sigma)}|^2,
 \\
 W_{\lambda}^{(m,\rm em)} &=& \sum\limits_{\sigma} \left[\left(\frac{1}{2} + 2\sigma\lambda\cos\theta\right)|{\bB}^{(\sigma)}|^2 -
 (\sigma + \lambda \cos\theta)\, \text{Im} \left(\bB^{(\sigma)*}\times{\bB}^{(\sigma)}\right)_z\right]
  \nn
   \\
 &-& 2\lambda\sin\theta \left[\mbox{Im}\left(\bB^{(1/2)*}\times\bB^{(-1/2)}\right)_x -
  \mbox{Re}\left(\bB^{(1/2)*}\times\bB^{(-1/2)}\right)_y \right],
  \\
 W_{\lambda}^{(m,\,\rm int)} &=&-\sum\limits_{\sigma} (\sigma + \lambda \cos\theta)\, \text{Im} I_z^{\sigma \sigma}-
 2\lambda\sin\theta \left[\mbox{Im} I_x^{(1/2,-1/2)} -
  \mbox{Re} I_y^{(1/2,-1/2)} \right].
   \eea

Note the following features:

1. This angular distribution contains the evident dependence on helicity
$\lambda$ of the initial neutron.

\begin{figure}
\centering
\includegraphics[scale=0.3]{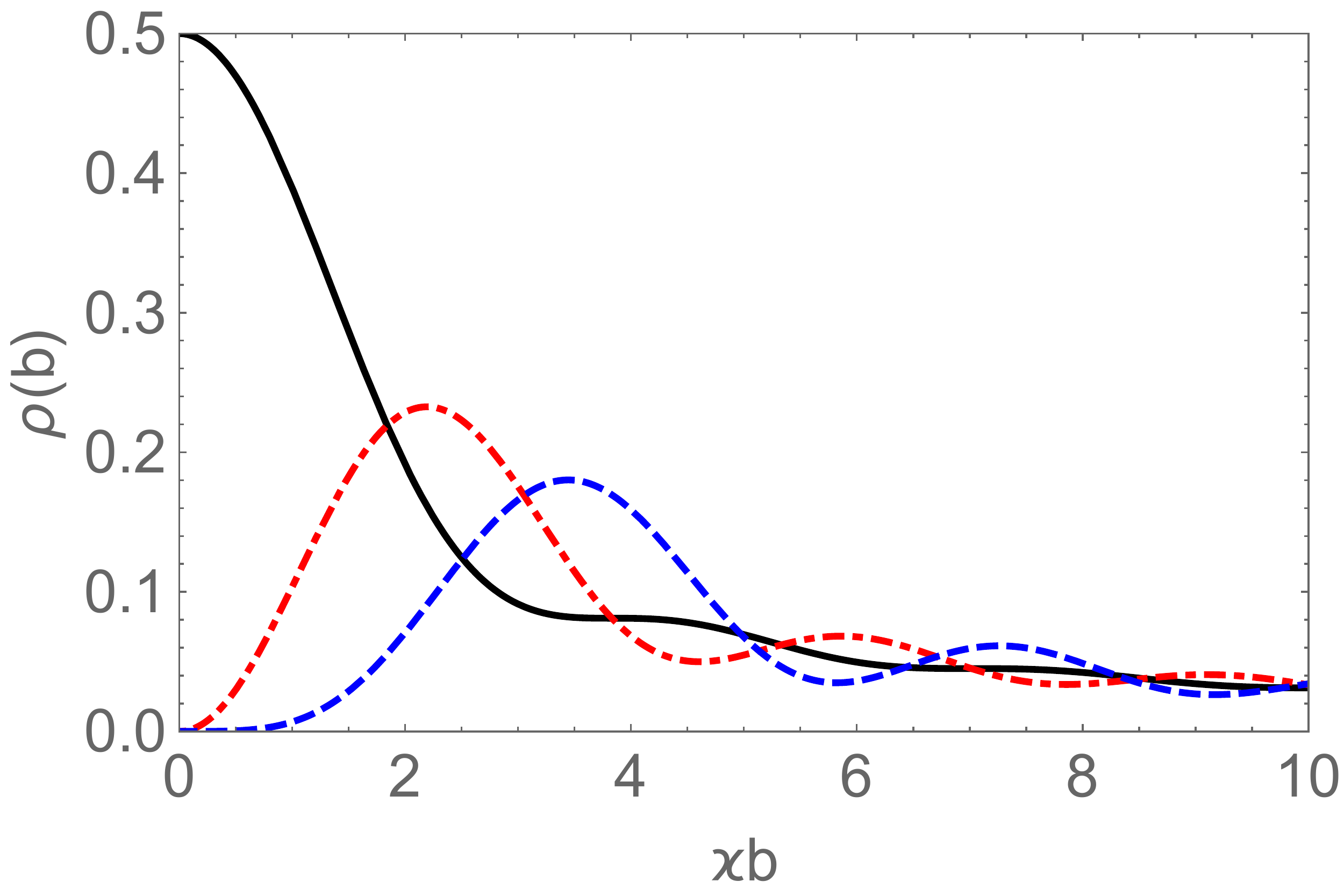}
\caption{The functions $\rho(b)$ from Eq.~\eqref{rho} for $m=1/2$ (solid black line), $m=3/2$ (red dot-dashed line)
and $m=5/2$ (blue dashed line) }
\label{Fig3old}
\end{figure}

2. The pure nuclear contribution is directly proportional to the density of incoming neutrons
(see Eq.~\eqref{Flux}):
 \be
W_\lambda^{(m,\,\rm nucl)}= |a|^2 \rho^{\,(m \lambda)}(\bb)=|a|^2 \rho (b)+ \lambda\, |a|^2
\cos\theta\, \left[J^2_{m-1/2}(\varkappa b)- J^2_{m+1/2}(\varkappa b)\right],
  \label{nuclear-contributin}
  \ee
where $\rho(b)$ is the density of neutrons averaged over their helicities:
 \be
 \rho (b) = \fr 12 \,\sum_{\lambda} \rho^{\,(m \lambda)}(\bb)= \fr 12 \, \left[J^2_{m-1/2}(\varkappa b)+ J^2_{m+1/2}(\varkappa b) \right]
 \label{rho}
 \ee
This function is shown in Fig.~\ref{Fig3old} for different values of $m$.

3. The interference of nuclear and electromagnetic interactions of neutrons is described by the terms ${\bf I}^{(\tilde\sigma\sigma)}$ only.

\section{Appendix B. The limit $b=0$}

In the limit of zero impact parameter $\bb\to 0$, the expressions~\eqref{Asigma} and
 \eqref{Bsigma}  for $A^{(\sigma)}$ and $\bB^{(\sigma)}$ are simplified. Indeed, using the formulas (27)--(28) from~\cite{Mott-2015}:
 \be
 \int_0^{2\pi} \fr{d\varphi}{2\pi} \fr{\bee^{\bii n \varphi}}{1-cc' - ss' \cos\varphi}=
 \fr{1}{|c-c'|}\,
 \left(\fr{ss'}{1-cc'+ |c-c'|} \right)^{|n|},
 \ee
we get
 \bea
   \label{Batb=0}
  A^{(\sigma)} &=& a\, \delta_{\sigma\, m},
   \nn
   \\
   \bB^{(\sigma)} &=& B_{m-\sigma}\;
 \left(\pm\bii c',\, \fr{c-c'}{|c-c'|},\, \mp\bii s'\right)\;\;\mbox{for}\;\;m-\sigma \gtrless 0,
  \\
  \bB^{(\sigma)} &=& B_{0}\;
 \left(0,\, \fr{c-c'}{|c-c'|},\, 0 \right)\;\;\mbox{for}\;\;m-\sigma= 0,
  \nn
    \eea
where
 \be
 B_{m-\sigma}=\fr{\beta}{2s'}\,\left(\fr{ss'}{1-cc' +|c-c'|} \right)^{|m-\sigma|}.
 \ee

Sometimes it is useful to employ the identity
 \be
 1-cc'+ |c-c'|= (1\pm c)(1 \mp c')\;\; \mbox{for}\;\; \theta' \gtrless \theta
 \ee
and transform the  above equation to the form
 \be
 B_{m-\sigma}=\fr{\beta}{2s'}\,\left(\fr{\tan(\theta/2)}{\tan(\theta'/2)}\right)^{\pm |m-\sigma|}
 \;\; \mbox{for}\;\; \theta' \gtrless \theta.
 \ee

\subsubsection{The case $m\neq \pm 1/2$}

In this case the flux of neutrons~\eqref{Flux} at $b=0$ disappears and only the contribution of the long-range electromagnetic interaction survives:
 \be
 \label{AngularDistributionb=0}
 W^{(m)}_{\lambda}(\theta, \theta', \varphi', \bb=0)= 2\,\fr{1\pm c'}{1\pm c}
 \left(1\pm \Lambda \right)\, \left(\fr{\beta}{2s'} \fr{\tan(\theta/2)}{\tan\theta'/2)}\right)^{2|m|-1}\;\;\;\mbox{for}\;\; \theta' \gtrless \theta
  \ee
with the notation
 \be
 \Lambda = 2\lambda\, \mbox{sign}(m).
 \ee
This situation can be called as {\it the scattering in the dark} in analogy with {\it the excitation in the dark}, which was observed in the experiment~\cite{Mainz} with
twisted photons.

The helicity asymmetry in this limit has a simple analytical expression
 \be
A_{\lambda}(\theta, \theta',\varphi', \bb=0)=\fr{W^{(m)}_{\lambda=1/2}(\theta, \theta', \varphi', \bb=0)-W^{(m)}_{\lambda=-1/2}(\theta, \theta', \varphi', \bb=0)}
 {W^{(m)}_{\lambda=1/2}(\theta, \theta', \varphi', \bb=0)+W^{(m)}_{\lambda=-1/2}(\theta, \theta', \varphi', \bb=0)}
 = \fr{c-c'}{|c-c'|}\, \mbox{sign}(m).
 \label{A(b=0)true}
 \ee

In is interesting to note that another asymmetry defined as
 \be
A_{\rm m}(\theta, \theta', \bb=0)=\fr{W^{(m)}_{\lambda=1/2}(\theta, \theta', \varphi', \bb=0)-W^{(m-1)}_{\lambda=-1/2}(\theta, \theta', \varphi', \bb=0)}
 {W^{(m)}_{\lambda=1/2}(\theta, \theta', \varphi', \bb=0)+W^{(m-1)}_{\lambda=-1/2}(\theta, \theta', \varphi', \bb=0)}
 \label{A(b=0)}
 \ee
has the same behavior for the case $m\neq \pm 1/2$ and $m-1\neq \pm 1/2$.

This asymmetry is relevant to specific experimental conditions, since it corresponds to a difference in the scattering rates between the neutrons of opposite spins $\lambda$, but with a fixed value of orbital angular momentum projection (=$m-\lambda$ in a paraxial approximation). Thus, it can be measured by only manipulating the spin degree of freedom $\lambda$ for the given twisted-neutron beam.

\subsubsection{The case $m=\pm 1/2$}

In this case the flux of neutrons~\eqref{Flux}  does not disappear at $b=0$ and all contributions do survive:
 \be
  W^{(m)}_{\lambda}(\theta, \theta', \varphi', \bb=0)= \fr 12 (\Sigma + \Lambda \Delta),
 \ee
where
 \bea
 \Sigma &=& |a|^2+ \left(\fr{\beta}{2s'}\right)^2\,\left[1+2H(\theta, \theta') \right],\;\;
 H(\theta, \theta')=\fr{(1\mp c)(1\pm c')}{1\pm c}\\
 \Delta &=& |a|^2 c \pm \beta \,\mbox{Re}(a)\,(1\mp c)+
 \left(\fr{\beta}{2s'}\right)^2\,\left[c \pm 2H(\theta, \theta') \right]
  \eea
Here, the upper (lower) sign corresponds to $\theta' > \theta$ ($\theta' < \theta$).

%

\end{document}